\documentclass[twocolumn]{jpsj2}

\def\runauthor{G. Misguich and M. Oshikawa}
\newcommand{\mue}{\mu^{\rm eff}}
\newcommand{\intdk}{\int \frac{d^3k}{(2\pi)^3}}
\newcommand{\et}{\tilde\epsilon_k}

\author{Gr\'egoire {\sc Misguich}$^{1}$ and Masaki {\sc Oshikawa}$^{2}$}
\inst{
$^{1}$Service de Physique Th\'eorique,
CEA-Saclay, 91191 Gif-sur-Yvette C\'edex, France\\
$^{2}$Department of Physics, Tokyo Institute of Technology, 
Oh-oka-yama, Meguro-ku, Tokyo 152-8551 Japan
}

\title{
Bose-Einstein Condensation of Magnons in TlCuCl$_3$: Phase diagram and
specific  heat from a   self-consistent Hartee-Fock calculation with a
realistic dispersion relation.}

\abst{
We  extend  the  self-consistent Hartree-Fock-Popov  calculations   by
Nikuni   {\it et  al.}  [Phys.   Rev.   Lett.   {\bf 84}, 5868 (2000)]
concerning the  Bose-Eistein condensation of  magnons in TlCuCl$_3$ to
include a realistic dispersion of the excitations.  The result for the
critical   field as     a   function  of   temperature    behaves   as
$H_c(T)-H_c(0)\sim T^{3/2}$ below   2~K but deviates from this  simple
power-law at higher temperature and is in very good agreement with the
experimental results.  The specific heat  is computed as a function of
temperature for different  values of the  magnetic field.  It  shows a
$\lambda$-like shape  at the transition   and is in  good  qualitative
agreement with  the results of Oosawa  {\it et  al.}   [Phys.  Rev.  B
{\bf 63}, 134416 (2001)].}

\begin{document}
\maketitle

\section{Introduction}

TlCuCl$_3$   is a spin-$\frac{1}{2}$  magnetic  insulator  with a spin
gap\cite{takatsu97}  of  $\Delta=7.5$~K\cite{shiramura97}. It has been
successfully   described as    copper   dimers  with   an  intra-dimer
antiferromagnetic exchange     energy  $J\simeq5.5$meV  and     weaker
($\lesssim1.5$meV) inter dimer couplings.\cite{cavadini01,matsumoto02}
% MO: added more explanation. ``closing of the gap'' is not
% quite well-defined at finite temperature??
At zero temperature, an applied magnetic field $H$ closes the
gap at the critical field $H=H_c(0)$, giving
rise to a quantum phase transition.
$H_c(0)$ is related to the gap by $g \mu_B H_c(0)=\Delta$,
where $\mu_B$ is the Bohr magneton and $g$ is the Lande $g$-factor.
The field-induced phase transition continues to finite temperature $T$,
with the temperature-dependent critical field $H_c(T)$.
Above the critical field, a magnetic long-ranged order in
the         plane    perpendicular      to      applied          field
develops.\cite{oit99,tanaka01}
% end MO
The     existence   of    the  ordering
transition was     predicted  by a    standard   mean-field theory for
spins\cite{TY}.   However,  several  characteristic features   of  the
transition could not be explained  by the mean-field theory.  The  two
most notable  features are the  cusp-like minimum of the magnetization
as a function of the temperature at the  transition, and the power-law
like dependence of the critical field
\begin{equation}
H_c(T) - H_c(0) \propto T^{\phi}
\label{eq:powerlaw}
\end{equation}
in the low temperature regime. 
The mean-field theory\cite{TY} rather predicts a monotonic decrease
of the magnetization and an exponentially fast approach of the
critical field $H_c(T)$ to its zero-temperature limit $H_c(0)$,
on lowering the temperature.

These features were successfully explained, at least qualitatively, as
a  Bose-Einstein  condensation   (BEC)  of  spin  triplet  excitations
(magnons).\cite{noot00} The cusp-like  minimum of the magnetization at
the  transition  temperature is understood with  the  decrease  of the
non-condensed  magnons at all  temperatures  and the  increase  of the
condensed magnons below the transition, as the temperature is lowered.
Moreover, the  self-consistent Hartree-Fock-Popov  (HFP) approximation
on   the       magnon   condensation   gives         the     power-law
dependence~(\ref{eq:powerlaw})  with the  exponent  $\phi=3/2$, if the
dispersion of the magnons is taken to be quadratic.

As one can easily control the magnetic field, which corresponds to the
chemical potential  of the magnons, this system  provides a  new arena
for  the study of  BEC, in a  grand-canonical  ensemble with a tunable
chemical potential.
\cite{rice02}

However,  the   results    of  the   HFP   approximation   given    in
Ref.~\citen{noot00}  are   not  quite  satisfactory  to   describe the
experimental data in a quantitative manner. In order to further extend
the study  of  magnon BEC, it would   be important to improve the  HFP
approximation and clarify its range of validity.

One of  the   problems  is that   the HFP    approximation  predicts a
discontinuous jump of the magnetization at the transition temperature,
which is  not observed.  This is considered  to be  an artifact of the
HFP   approximation, and related  to     its breakdown due to   strong
fluctuation  in  the vicinity of  the transition.    In this  paper we
rather  focus on another problem  concerning the phase boundary.  That
is,  while the experimental results  are roughly  in agreement with
the     power     law~(\ref{eq:powerlaw}),           the      reported
values\cite{oit99,noot00,okt01,tanaka01,st03} of  the exponent $\phi =
1.67 \sim 2.2$ are consistently  larger than the HFP prediction $3/2$.
Although  it was suggested  that  the deviation is   again due to  the
fluctuation effects, it has not been clarified.

In    the  present    work,  we   extend   the    self-consistent  HFP
calculations\cite{noot00}  by   including    a realistic    dispersion
calculated    from    microscopic  models\cite{cavadini01,matsumoto02}
instead        of        the           quadratic         approximation
$\epsilon_k\simeq\frac{k^2}{2m}$    used previously.\cite{noot00}  The
critical  field  $H_c(T)$  obtained by  this method   is in  very good
agreement    with the   experiments    and  represents  a  significant
improvement over  the simple quadratic  approximation.  Therefore  the
puzzle regarding the discrepancy   of the exponent $\phi$  between the
theory and the experiment is solved within the HFP framework.  Here we
note that there are related theoretical
% GM: Ref. to Sirker added in the list
works\cite{sherman03,normand04,nohadani03,sirker04}  on  this problem.
We will comment on them later in Discussions.

We  also make several other  checks of the  HFP approximation with the
experimental data, to show that HFP  framework has a rather wide range
of validity but the quadratic  approximation fails above a rather  low
temperature $\sim 1$ K for  TlCuCl$_3$.  Finally, the specific heat is
also  computed and    compared with  the  results  of  Oosawa  {\it et
al.}\cite{okt01}

\section{Hamiltonian}

As  in Ref.~\citen{noot00}, the   Zeeman  splitting is  assumed  to  be
sufficiently large compared to temperature so that only the singlet and
the lowest triplet states  of each dimer  need to be considered.  With
this approximation the system  is   described by an  hard-core   boson
Hamiltonian
\begin{equation}
        \mathcal{H}=\mathcal{H}_K+\mathcal{H}_U
        \label{eq:H}
\end{equation}
$\mathcal{H}_K$ contains   the   zero-temperature  magnon   dispersion
relation $\epsilon_k+\Delta$ and the external magnetic field $H$:
\begin{eqnarray}
        \mathcal{H}_K&=& \sum_k b^\dagger_k \; b_k \left(\epsilon_k-\mu\right)\\
        \label{eq:HK}
        \mu&=&g\mu_B H-\Delta
\end{eqnarray}
where   it  is    assumed  that   $\epsilon_0=0$.  The   magnon-magnon
interactions are described by
\begin{equation}
        \mathcal{H}_U= \frac{1}{2N} \sum_{q,k,k'}
                U_q \;\;
                b^\dagger_k \; b^\dagger_{k'} \; b_{k+q} \;b_{k'-q} 
        \label{eq:HU}
\end{equation}
and we will neglect the $q$-dependence of $U_q$ and set $U_q=U$.  As
discussed by Nikuni  {\it et al.}~\cite{noot00}, this system undergoes
a phase  transition  between a  normal  phase  (at low field  or  high
temperature) where the    system is  populated  by thermally   excited
triplets to  a ``superfluid'' phase where  the  bosons condense.  This
condensation is equivalent, in  the spin language, to a  field-induced
three-dimensional magnetic ordering.

%______________________________________________________________________
\section{Hartree-Fock-Popov treatment of the condensed phase}

We   reproduce  the Hartree-Fock-Popov   (HFP)  mean-field analysis of
Eqs.~\ref{eq:H}-\ref{eq:HU}        which         was   discussed    in
Ref.~\citen{noot00}.    For a    strong   enough   magnetic  field  the
zero-momentum state (we assume that  $\epsilon_k$ has a single minimum
at      $k=0$)        is            macroscopically           occupied
$b^\dagger_{k=0}=b_{k=0}=\sqrt{N_c}=\sqrt{N   n_c}$.  From this we can
write the interaction part of the  Hamiltonian in terms of a constant,
2-, 3- and 4-boson operators:\cite{pn66}
\begin{eqnarray}
        \mathcal{H}_U&=& H_0 + H_2 + H_3 + H_4  \\
        \mathcal{H}_0&=& \frac{1}{2N} \; U \; N_c^2 \\
        \mathcal{H}_2&=&\hspace*{-0.2cm}  \frac{UN_c}{N} \sum_q{'} \left[
         \;\frac{1}{2}\left( b_q \;b_{-q}+b^\dagger_{-q} \; b^\dagger_q\right)
                + 2b^\dagger_q\;b_q 
        \right] \\
        \mathcal{H}_3&=& U\frac{\sqrt{N_c}}{N} \sum_{k,q}{'}\left(
        b^\dagger_k \; b_{k+q} \; b_{-q}
        + {\rm H.c} \right) \\
        \mathcal{H}_4&=& \frac{U}{2N} \sum_{q,k,k'}{'}
                b^\dagger_k \; b^\dagger_{k'} \; b_{k+q} \;b_{k'-q} 
\end{eqnarray}
where   $\sum'$ means  that the  terms   with creation or annihilation
operators   at $k=0$ are  excluded.    We perform  a simple mean-field
decoupling for $\mathcal{H}_U$. While $\mathcal{H}_3$ gives
zero in this approximation, $\mathcal{H}_4$ gives:
\begin{equation}
        \mathcal{H}_4^{\rm MF}=- U_0 N (n-n_c)^2
        +2 (n-n_c) U_0 \sum_k{'} b^\dagger_k \;  b_k
\end{equation}
% GM: 'n' was missing in the sentence below
where $n$ is the     total  boson   density;   it   must  be    determined
self-consistently  from  the  thermal  average over   the  spectrum of
$\mathcal{H}^{\rm MF}=\mathcal{H}_0+\mathcal{H}_2+\mathcal{H}_4^{\rm MF}$:
\begin{eqnarray}
        \mathcal{H}^{\rm MF}&=&C+ \sum_k{'}
                \et \;b^\dagger_k \;  b_k \nonumber\\
                &&+\frac{Un_c}{2}\;\sum_q{'} 
        \left( b_q \;b_{-q}+b^\dagger_{-q} \; b^\dagger_q\right)
        \label{eq:hmf}\\
        \et&=&\epsilon_k -\mue \;\;,\;\;\mue=\mu - 2 Un\label{et}\\
        C&=&U N \left[ \frac{1}{2}n_c^2-(n-n_c)^2\right] -\mu n_c
\end{eqnarray}
The mean-field Hamiltonian in the normal phase  is obtained by setting
$n_c=0$ in the previous expression (already in a diagonal form). In that case,
the self-consistent equation for the density is
\begin{equation}
  n=\intdk f_B(\et)\label{eq:scn}
\end{equation}
where $f_B(E)=1/(\exp(\beta E)-1)$ is the Bose occupation number.

When $n_c>0$,    $\mathcal{H}^{\rm MF}$ can    be  diagonalized by the
standard Bogoliubov transformation:
\begin{eqnarray}
  \mathcal{H}^{\rm MF}&=&\hspace*{-0.2cm}\sum_k{'} E_k \; (\alpha^\dagger_k \; \alpha_k + \frac{1}{2})
  - \frac{1}{2}\sum_k{'} \et
  +C
  \label{HMF}
  \\
  E_k&=&\sqrt{\et\;^2-\left(U\;n_c\right)^2}
  \label{Ek}
  \\
  b_k&=& u_k \;\alpha_k - v_k \;\alpha^\dagger_{-k}
  \label{alpha}
  \\
  u_k&=&\sqrt{\frac{\et}{2E_k}+\frac{1}{2}}
  \;\;,\;\;
  v_k=\sqrt{\frac{\et}{2E_k}-\frac{1}{2}}
  \label{uv}
\end{eqnarray}
The existence of  a  condensate ($n_c>0$)  is  possible when $E_k$  is
gapless, which implies $\mue=-Un_c$, or equivalently:
\begin{equation}
        g\mu_BH=\Delta+U\left(2n-n_c\right) \label{mu}
\end{equation}
$\mue$, $n$ and $n_c$ are thus linearly related in the condensed phase
and the self-consistent equation is now:\cite{noot00}
\begin{equation}
  n-n_c=\intdk \left[
    \frac{\et} {E_k} \left(f_B(E_k)+1\right)
  \right]-\frac{1}{2}
  \label{eq:scc}
\end{equation}

\section{Dispersion relation for TlCuCl$_3$}

\begin{figure}
\begin{center}
	% GM: Fig. 'disp.eps' was corrected: 5.489 K -> 5.489 meV	
	% \epsilon\simeq k^2/(2m) replaced by ``Quad approx.''
        \includegraphics[width=7cm]{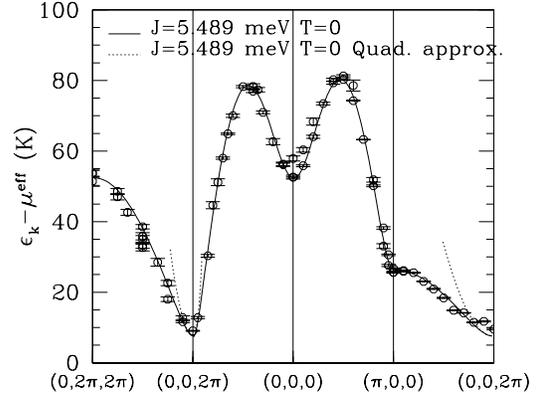}
        \end{center}
\caption[99]{Dispersion relation of triplet excitations.
Full lines: Result
of Eq.~\ref{eq:ek} with $J$ given by Eq.~\ref{eq:J}.
% -----
% GM:
Dotted line: (anisotropic) quadratic approximation in the vicinity of the minimum.
% -----
Circles and error bars are from Ref.~\citen{cavadini01}.
% -----
% GM:
The labels of the horizontal axis represent $k'=k+k_0$ 
to reconcile the convention $\epsilon_{k=0}=0$ and the location of the minimum
of the triplet dispersion in  TlCuCl$_3$ at  momentum $k'=k_0=(0,0,2\pi)$.
% -----
}
\label{fig:disp}
\end{figure}

The dispersion  relation  of  triplet  excitations  in TlCuCl$_3$  was
measured at  $T=1.5$~K with inelastic  neutrons scattering by Cavadini
{\it  et al.}\cite{cavadini01}  This  dispersion relation  was very well
reproduced   by Matsumoto    {\it et al.}\cite{matsumoto02}   within a
bond-operator formalism. Their result is:
\begin{eqnarray}
        \epsilon_{k-k_0}+\Delta_0&=&\sqrt{(J+a_k)^2-a_k^2}
        \label{eq:ek}\\
        a_k&=&J_a\cos(k_x)+J_{a2c}\cos(2k_x+k_z)\nonumber\\
                &&\hspace*{-0.4cm}+2J_{abc}\cos(k_x+k_z/2)\cos(k_y/2) \\
        J&=&\hspace*{-0.2cm}5.501\;{\rm meV} \;,\;\; J_a=-0.215\;{\rm meV} \\
	% GM: sign of J_{a2c} corrected
        J_{a2c}&=&\hspace*{-0.2cm}-1.581\;{\rm meV} ,\;J_{abc}=0.455\;{\rm meV}
\end{eqnarray}
where the Brillouin zone is doubled in  the $z$ direction ($-2\pi\leq
k_z<2\pi$) to   represent the  two  magnon branches.
% -----
% GM
The momentum shift by $k_0=(0,0,2\pi)$ just insures the consistency between
our convention that $\epsilon_{0}=0$  and the location of  the minimum
of the dispersion at $k_0$ in Refs.~\citen{cavadini01,matsumoto02}
% -----
The  dispersion
relation above has a gap of $0.7$~meV, which  is in agreement with the
result  of Ref.~\citen{cavadini01}.   However  the studies  based on  a
determination  critical   field  as a   function  of  temperature (see
Table~\ref{tab:gap}) provide slightly  smaller  estimates for the  gap
($\Delta_0\sim 0.65$~meV) in TlCuCl$_3$.   Therefore we  corrected the
value of $J$ so that the dispersion  relation is consistent with these
data. The  corrected value was  chosen to  insure $\Delta_0=0.65$~meV
(or equivalently $(g/2)H_c(0)=5.61$T) :
\begin{equation}
        J=5.489\;{\rm meV}
        \label{eq:J}
\end{equation}
From the computation  of  curvature of  $\epsilon_k$ around $k=0$  the
effective inverse mass\cite{mass}  $1/m$ is $43.66$~K (in units  where
$\hbar^2/k_B=1$),    in    agreement   with   the   value   taken   in
Ref.~\citen{noot00}. Fig.~\ref{fig:disp} shows the experimental   data
of  Cavadini {\it    et     al.}  with the   $\epsilon_k$    given  by
Eqs.~\ref{eq:ek}-\ref{eq:J}.   The  dotted  line  corresponds  to  the
quadratic approximation;  it only matches the  full expression at very
low energy.

\begin{table}[tb]
\caption{
  Estimations of the gap (or  critical field at zero temperature) from
  experiments.   Mag.   stands for  magnetization,   INS  for inelastic
  neutron scattering, ESR for electron spin resonance, ENS for elastic
  neutron scattering (observation  of the magnetic ordering) and $C_v$
  for specific heat measurements.}\vspace{0.5cm}
\label{tab:gap}
\begin{tabular}{|ccc|}
  \hline
  Ref. & $\Delta_0$ & Method \\
  \hline
  Shiramura {\it et al.}\cite{shiramura97}& 7.5~K         & Mag. \\
  (1997)&($\frac{g}{2}H_c=5.6$T)&\\
  \hline
  Tanaka {\it et al.}\cite{ttts98}           & 7.68~K        & ESR           \\
  (1998)& (160 GHz)     &               \\
  \hline
  Oosawa {\it et al.}\cite{oit99}            & 7.54 K        & Mag. \\
  (1999)&($\frac{g}{2}H_c=5.61$T)&\\
  \hline
  Tanaka {\it et al.}\cite{tanaka01}         & 7.66 K        & ENS           \\
  (2001)& ($\frac{g}{2}H_c=5.7$T) &\\
  \hline
  Cavadini {\it et al.}\cite{cavadini01}     & 9.28K         & INS           \\
  (2001)& (0.8meV)      &               \\
  \hline
  Oosawa {\it et al.}\cite{okt01}            & 7.66K         & $C_v$         \\
  (2001)& ($\frac{g}{2}H_c=5.7$T)&\\            
  \hline
  Oosawa {\it et al.}\cite{okt02}            & 7.54K         & INS           \\
  (2002)& (0.65meV)     & \\
  \hline
  R\"uegg {\it et al.}\cite{ruegg03}          & 8.2K          & INS           \\
% GM: Value of the gap modified : 0.75meV -> 0.71meV(8.2K) (according to Ruegg's e-mail)
  (2003)& (0.71meV)     & \\
  \hline
  Shindo {\it et al.}\cite{st03}             & 7.33K         & $C_v$         \\
  (2003)& ($\frac{g}{2}H_c=5.46$T)&\\
  \hline
\end{tabular}
\end{table}

\section{Critical density}

\begin{figure}
\begin{center}
        \includegraphics[width=7cm]{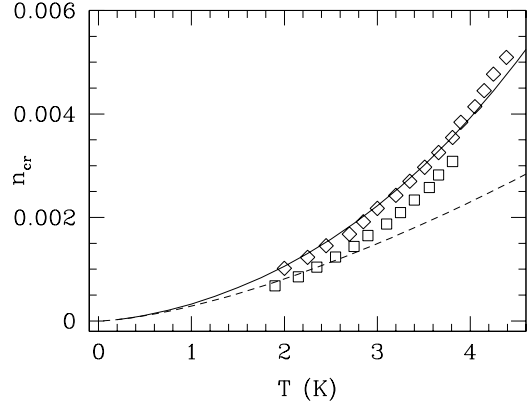}
        \end{center}
\caption[99]{Critical boson density as a function of temperature.
Squares:  magnetic field  along the $b$   direction.  Tilted squares :
magnetic   field   along  the    $(1,0,\bar2)$   direction (data  from
Oosawa~{\it et al.}\cite{oit99}). Full line: HFP  result with the full
dispersion  relation.   Dotted line:   HFP result  with  the quadratic
approximation            for      the   dispersion            relation
$\epsilon_k=k^2/(2m)$ and $k_B/m=43.6$~K.}
\label{fig:n}
\end{figure}

Within the HFP   approximation  the boson  density  $n_{\rm   cr}$ (or
magnetization) at the transition is independent of the strength $U$ of
the magnon-magnon interaction  as well as independent  of the value of
the zero-field gap $\Delta_0$.  It  is obtained by setting $\mue=0$ in
Eq.~\ref{eq:scn}.\cite{remark1} If  the  full dispersion  relation  is
used, the result has {\em no adjustable parameter} left.  The result is
shown Fig.~\ref{fig:n} and is in good  agreement with the experimental
data.   We note however that the  discrepancy is larger when the field
is applied along  the   $b$ direction.
% MO: clarified that we don't understand this.
We do not know the reason of the discrepancy at present.
% also, we'd better move to a new paragraph here.
% end MO 

In the low-temperature limit, the quadratic approximation
would become asymptotically exact within the HFP theory,
giving~\cite{noot00,remark2}
\begin{equation}
        n_{\rm cr}(T\to 0)= \frac{1}{2}\zeta_{3/2} \left(\frac{T m}{2\pi}\right)^{3/2}
        \label{eq:zeta}
\end{equation}
However,  this     $\sim   T^{3/2}$   behavior   (dotted    line    in
Fig.~\ref{fig:n})   is only recovered    at  very low temperature  and
$n_{\rm cr}(T)$  shows  significant deviations  from Eq.~\ref{eq:zeta}
already at 2~K.

\section{Critical field and interaction parameter $U$}

\begin{figure}
\begin{center}
	\includegraphics[width=7cm]{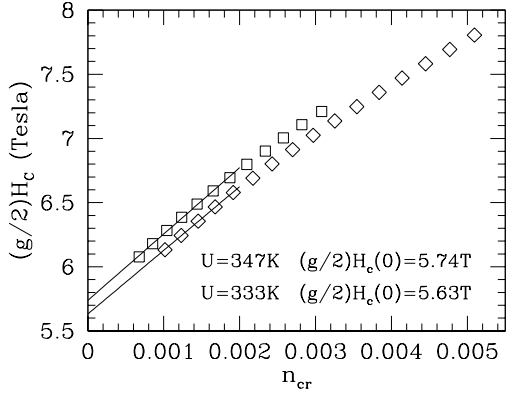}
\end{center}
\caption[99]{
Critical  field $H_c$ (normalized   by  the $g$  factor  and for   two
magnetic  field direction: squares for $H||b$  and $g=2.06$ and tilted
squares for $H\perp(1,0,\bar{2})$ and $g=2.23$)  as a function of  the
density $n_{\rm cr}$ at the    critical point (obtained from the   the
magnetization  $m_{\rm    cr}$  per   dimer   by   $n_{\rm  cr}=m_{\rm
cr}/(g\mu_B)$).  Data from Ref.~\citen{oit99}.  The full lines and the
values     of  $U$  and $H_c(0)$       are   obtained  from  fits   to
Eq.~\ref{eq:hcUnc}.}
\label{fig:hcnc}
\end{figure}

\begin{figure}
\begin{center}
	\includegraphics[width=7.5cm]{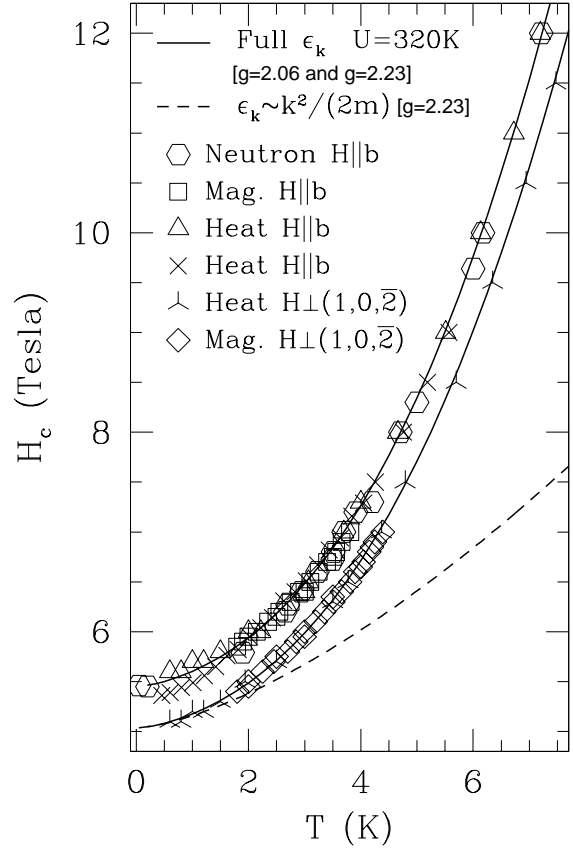}
\caption[99]{
Critical  field  $H_c(T)$.  Full    line:  HFP result with  the   full
dispersion  relation  and $U=320$~K  (plotted  for  two  values of the
gyromagnetic factor).   Dotted     line:  $\epsilon_k=k^2/(2m)$
approximation ($g=2.23$).   Hexagons: Data from Ref.~\citen{tanaka01}.
Squares and crosses  :  Data from Ref.~\citen{okt01}.   Triangles  and
three-leg symbol: Data from Ref.~\citen{st03}.}
\label{fig:h}
\end{center}
\end{figure}

In the HFP approximation the critical field $H_c(T)$ is related to the
critical density  by~\cite{noot00}
\begin{equation}
   (g/2)\left[H_c(T)-H_c(0)\right]=2 U  n_{\rm cr}(T)
        \label{eq:hcUnc}
\end{equation}
A linear  relation between   $H_c(T)$  and $n_{\rm cr}(T)$   is indeed
observed  in     the  experimental  data,   as   can   be   seen    in
Fig.~\ref{fig:hcnc}.   The least-square  fits  are  performed in  the
low-density   region (or  equivalently  low-temperature).  The  values
obtained for $H_c(0)$ are in good agreement  with most of the previous
estimates  (see Tab.~\ref{tab:gap}).   These    fits also  provide  an
estimate  for $U$  around  $340$~K.  However, as  it  can  bee seen in
Fig.~\ref{fig:h}, a  slightly smaller value  for $U$ ($320$~K) gives a
critical field $H_c(T)$  which is in very  good agreement with all the
available experimental  data,  even  at  high temperatures. 
% ------
% GM: comment added
This value   is close to that  obtained  from a similar   HFP analysis
(including a small  magnetic exchange anisotropy) of the magnetization
curves.\cite{sirker04}
% ------

In  the  literature the  experimental  data  for  $H_c(T)$  have been
analyzed by fitting to the power-law~(\ref{eq:powerlaw}).
Values     from      $\phi=1.67$      to      $2.2$       have    been
reported\cite{oit99,noot00,okt01,tanaka01,st03}     and   it  has been
suggested that the deviation from the HFP theory ($\phi=1.5$) could be
caused by fluctuations   effects beyond the mean-field  approximation.
% MO: clarified what ``a wide temperature range'' means
From    our  results   it appears    that      a realistic  dispersion
relation\cite{matsumoto02} combined with  an HFP treatment is  able to
reproduce the  data accurately with a single adjustable parameter ($U$).
It covers a wide temperature range from the very low temperature
regime $< 1$ K where the quadratic approximation holds, up to
$\sim 8$ K. 
% end MO

%______________________________________________________________________
\section{Specific heat}
\label{sec:C}

The  specific heat of TlCuCl$_3$ under  magnetic field was measured by
Oosawa  {\it et al.}~\cite{okt01} and shows  a peak (with an asymmetric
$\lambda$ shape)  at the transition. In this  section we compare these
results with the prediction of the HFP theory.

%______________________________________________________________________

From Eq.~\ref{eq:hmf} the expectation value of the  energy per site in
the normal phase ($n_c=0$) is 
\begin{equation}
        \langle E \rangle=-U n^2+\intdk \et f_B(\et,T)
\end{equation}
The  specific  heat is  obtained  by  differentiation with  respect to
temperature and we get:
\begin{eqnarray}
        C_v=
        \frac{1}{T} \intdk \et^2\left(-\frac{\partial f_B}{\partial \et}\right)
        \nonumber \\
        +2U\frac{\partial n}{\partial T}
                \intdk \et\frac{\partial f_B}{\partial \et}
\end{eqnarray}
with $k_B=1$ and
\begin{equation}
        \frac{\partial n}{\partial T}=\frac{
        \intdk \frac{\partial f_B}{\partial T}
                                        }{
        1-2 U \intdk \frac{\partial f_B}{\partial \et}
        }                                       
\end{equation}
%______________________________________________________________________

In the condensed phase, Eq.~\ref{eq:hmf} gives
\begin{equation}
        \langle E \rangle=\intdk E_k \; \left[f_B(E_k,T)+ \frac{1}{2}\right]
        - \frac{1}{2}\intdk \et + C 
\end{equation}
After some algebra, we obtain the specific heat as:
\begin{eqnarray}
        C_v&=&
        \frac{1}{T} \intdk E_k^2\left(-\frac{\partial f_B}{\partial E}\right)
        \nonumber \\
        &&\hspace*{-0.3cm}+2U\frac{\partial n}{\partial T}
        \left[
                n_c-n-\frac{1}{2} \right. \nonumber\\
              && \hspace*{-0.3cm}\left.
                +\intdk \frac{\epsilon_k}{E_k}\left(
                        f_B(E)+\frac{1}{2}+E \frac{\partial f_B}{\partial E}
                        \right)
        \right]
\end{eqnarray}
with 
\begin{equation}
        \frac{\partial n}{\partial T}=\frac{
        \frac{1}{T}\intdk \et \frac{\partial f_B}{\partial E}
                                        }{
        1-2 U \intdk 
        \frac{\epsilon_k}{E_k^2}\left(
        -\et \frac{\partial f_B}{\partial E}+\frac{\mue}{E_k}(f_B+\frac{1}{2})
        \right)
        }                                       
\end{equation}

\begin{figure}
\begin{center}
	\includegraphics[width=7cm]{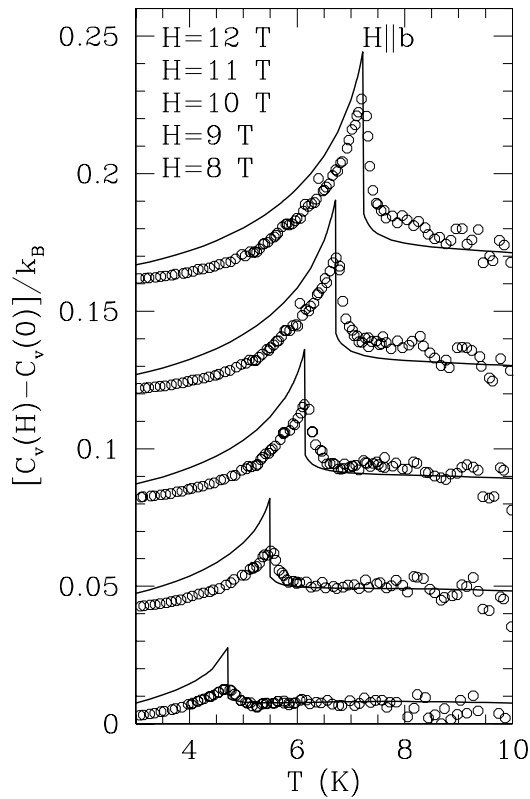}
        \end{center}
        \caption[99]{
          Specific heat (per dimer) under an  applied field (along the
          $b$ axis) minus the specific heat in zero field. Full lines:
          HFP results  with $U=320$K.  Circles: Measurements by Oosawa
          {\it et   al.}\cite{okt01}  The  results for  the  different
          values of $H$ have been shifted by 0.04 for clarity.}
\label{fig:cv}
\end{figure}
\begin{figure}
\begin{center}
	\includegraphics[width=7cm]{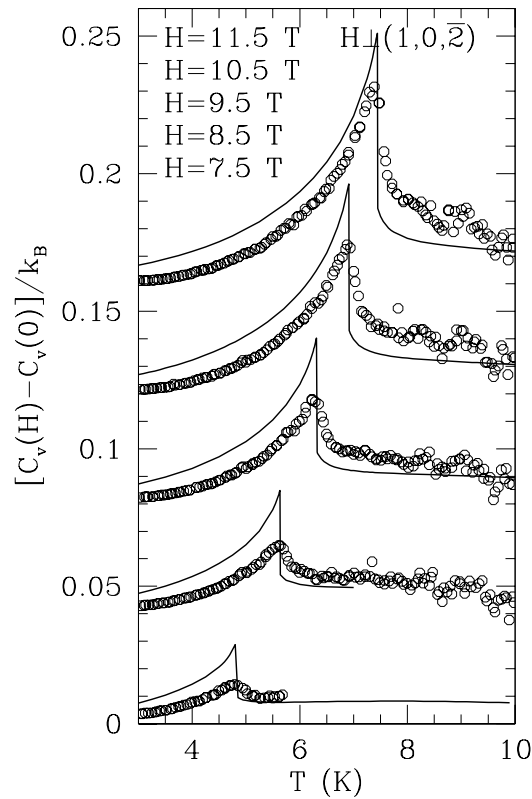}
	\end{center}
        \caption[99]{ Same  as Fig.~\ref{fig:cv}  with magnetic  field
        $H\perp(1,0,\bar2)$.}
\label{fig:cv2}
\end{figure}

In Figs.~\ref{fig:cv} and  \ref{fig:cv2}   the HFP results above   are
compared with the data of Oosawa {\it et al.}   for two magnetic field
orientations.   The  theoretical  curves  reproduce qualitatively  the
$\lambda$ shape  observed experimentally,  although the height  of the
peak seems to be overestimated.

\section{Discussions}

In this  paper, we  have  shown that  taking the  realistic dispersion
relation determined  from the microscopic  theory and from the neutron
scattering data, we can significantly improve the HFP approximation to
explain  the  experimental data, especially   the phase boundary curve
$H_c(T)$.  It is now evident that, in TlCuCl$_3$ the magnon dispersion
curve is rather ``steep'' so  that  the quadratic approximation  fails
above a rather low temperature $\sim 1$ K.

It  may be  rather  surprising that the  HFP  approximation,  which is
generally believed to fail in  the  critical region, describes a  wide
range  of experimental data  precisely.  This appears  to be the case,
even though  the  HFP   approximation still  contains   unsatisfactory
features of predicting discontinuities in the magnetization and in the
specific heat at the transition.  These discontinuities are considered
to be an  artifact of the HFP  approximation. The true behavior of the
magnetization in the model~(\ref{eq:H})  is believed to be  continuous
and  that of the specific heat
% -----
% GM: sentenced mofied
% to be logarithmically divergent at the
to   show  a   sharp    cusp    (negative   exponent   $\alpha$,   see
Ref.~\citen{campostrini01}) at the
% ----
transition, which is classified  as the 3-dimensional XY  universality
class.

However,   in fact, the  experimental data  on TlCuCl$_3$ discussed in
Section~\ref{sec:C} does not  show
% -----
% GM:
% the logarithmic divergence
such a sharp singularity 
% -----
and  is
rather similar to the HFP prediction.   This may be explained by small
anisotropies (breaking the $U(1)$  symmetry around the magnetic  field
direction), which are expected  to exist in  any real magnetic system.
The fact that the observed moment  in the ordered (condensed) phase of
TlCuCl$_3$ points to a  constant direction\cite{tanaka01} suggest  the
presence  of the anisotropy.  Moreover, recently it  is  argued that a
high-precision       ESR     measurement            reveals        the
anisotropy.\cite{kolezhuk04} Such anisotropies induce  a small gap and
should reduce the thermal fluctuations (and thus the specific heat) in
the vicinity of the transition, which could be also smeared out into a
crossover.  Since the breakdown of the  HFP approximation is generally
due to the critical fluctuation,  the reduction of the critical region
caused by the magnetic  anisotropies may  actually make the  agreement
with the   HFP  predictions  better,  although  we  did  not  take any
anisotropy  into  our   calculation.
% -----
% GM old sentence:
%    Further   improvement  could  be
%    possible by taking the anisotropy into the HFP calculation.
% new:
Recently, an HFP calculation  including a (small) magnetic  anisotropy
was carried out by Sirker {\it  et al.}\cite{sirker04} and provided an
improved  description of the   magnetization curves compared to that
obtained from the isotropic model.
% -----
% MO: a few more comment
They also emphasized that the HFP approximation should be valid
outside a narrow critical regime.
%  end MO

% GM: Added comment about the lattice distorsion
Magnetic anisotropies are not  the only corrections  that may be added
to  the present model.   Indeed,  NMR  measurements revealed that  the
transition  to  the ordered    phase   is (weakly) first   order   and
accompanied by a simultaneous lattice distortion\cite{vyaselev04} (see
also Ref.~\citen{sherman03}).  Spin-phonon interactions therefore  seem
to reduce the importance critical fluctuations close to the transition
while the resulting  lattice distortion certainly induces some  change
in the magnetic  exchange  parameters.\cite{ruegg} An analysis of  the
consequences of  such  a  magneto-elastic coupling is  an  interesting
issue for further studies.
% -----

Finally, let us comment on related theoretical works.  Sherman {\it et
al.} discussed that the agreement of  the HFP result to the experiment
is better if the ``relativistic'' form $\epsilon_k+\Delta\sim\sqrt{c^2
k^2   +  \Delta^2}$ is        assumed   for the     magnon  dispersion
relation.\cite{sherman03} Our approach in this  paper of modifying the
dispersion   is  actually the same   to theirs.   However,  we  see no
particular reason why we  should take the relativistic form,  although
it may be a  better approximation  for  TlCuCl$_3$ than the  quadratic
one.   In  any  case, ours  would  give   a  further improvement  over
Ref.~\citen{sherman03} within the HFP framework.

In Refs.~\citen{normand04,nohadani03}, the  phase boundary $H_c(T)$ is
studied numerically  by a Monte Carlo method,  for a dimer system on a
cubic    lattice.  The result should    contain  effects from both the
deviation of the dispersion from simple quadratic, and the fluctuation
beyond HFP While we cannot directly compare their result to ours as we
deal with  different  models, the   qualitative  behavior is  similar.
Namely,  they also observed the   deviation from $\phi=3/2$ at  higher
temperatures, but the result seems to  become closer to the $\phi=3/2$
as  the temperature is  lowered.   However,  they  suggest  that  this
behavior including the deviation   from $\phi=3/2$ could be  universal
and does not depend on the  particular dispersion, in a moderately low
temperature  regime.   This is in  contrast   to our  result that  the
non-universal  magnon dispersion explains  the observed phase boundary
$H_c(T)$ and its deviation from $\phi=3/2$.  The resolution is an open
problem for the future.  Numerical  approaches would be also useful to
clarify the effect of the (small) anisotropies.

\section{Acknowledgments}
We   thank  Hidekazu Tanaka  on  numerous stimulating  discussions and
providing  experimental   data  from many   experiments.
% MO: added those communicated to us
We also thank V.~N. Glazkov, N. Kawashima, 
Ch. R\"uegg and J. Sirker for useful correspondences, in particular
for notifying us of their work.
% end MO
G.~M.  and
M.~O. acknowledge the hospitality of the Tokyo Institute of Technology
and CEA Saclay respectively, on mutual visits.   M.~O. is supported in
part by Grant-in-Aid  for Scientific Research, and  a 21st Century COE
Program at Tokyo   Institute of  Technology  ``Nanometer-Scale Quantum
Physics'', both from MEXT of Japan.  G.~M. is in part supported by the
Minist\`ere de la Recherche et des  Nouvelles Technologies with an ACI
grant.     We thank  the KITP  for    hospitality during  part of this
work. This research  was  supported in part  by the   National Science
Foundation under Grant No. PHY99-07949 (KITP).

%MO: comment on Kawashima added
\section*{Note Added}
After submission of this paper,
Kawashima~\cite{Kawashima} clarified the question of the exponent $\phi$
with numerical simulations of the 3D S=1/2 XXZ model as well as
field-theoretical arguments. 
According to his results, in the limit of $T \rightarrow 0$,
the HFP prediction $\phi=3/2$ is indeed {\em exact}.
This is also consistent with our result that the phase boundary
for a wide temperature range can be accounted within the HFP
calculation using the realistic dispersion curve.
% end MO

% __________________________________________________________________________


\begin{thebibliography}{100}
\bibitem{mass} % Remark added (to answer the referee's questions)
The  dispersion relation  is  not isotropic  in   the vicinity of  its
minimum   but the  effective   mass    is  defined  by $m=(m_1     m_2
m_3)^{\frac{1}{3}}$ where  the $1/m_i$  are  the 3 eigenvalues  of the
quadratic form defined  by   $\epsilon_k$ around its  minimum.    When
$T\to0$, $m$ is the  only relevant combination of the  $m_i$ as far as
the total boson density is concerned (quadratic approximation).
\bibitem{remark2}
In the case of TlCuCl$_3$ there are {\em two}  magnon branches but the
minimum of the  dispersion relation (where  the effective mass $m$  is
defined) is  reached  in only {\em  one}.   This is the  origin of the
factor $\frac{1}{2}$ in Eq.~\ref{eq:zeta}.
\end{thebibliography}
\end{document}